\documentclass[a4paper,UKenglish]{lipics-v2016}

\usepackage{microtype}

\title{Improved Algorithms for Adaptive Compressed Sensing\footnote{This work was partially supported by NSF grant IIS-144741.}}
\titlerunning{Improved Algorithms for Adaptive Compressed Sensing} 

\author[1]{Vasileios Nakos}
\author[2]{Xiaofei Shi}
\author[3]{David P. Woodruff}
\author[4]{Hongyang Zhang}
\affil[1]{Harvard University, Cambridge, USA\\
  \texttt{vasileiosnakos@g.harvard.edu}}
\affil[2]{Carnegie Mellon University, Pittsburgh, USA\\
  \texttt{xiaofeis@andrew.cmu.edu}}
  \affil[3]{Carnegie Mellon University, Pittsburgh, USA\\
  \texttt{dwoodruf@cs.cmu.edu}}
  \affil[4]{Carnegie Mellon University, Pittsburgh, USA\\
  \texttt{hongyanz@cs.cmu.edu}}
\authorrunning{V. Nakos, X. Shi, D. P. Woodruff and H. Zhang} 

\Copyright{Vasileios Nakos, Xiaofei Shi, David P. Woodruff and Hongyang Zhang}

\EventLogo{eatcs}
\SeriesVolume{80}
\ArticleNo{}

\subjclass{F.2: Analysis of Algorithms and Problem Complexity}

\keywords{Compressed Sensing, Adaptivity, High-Dimensional Vectors}
\usepackage{algorithm,algorithmic}

\newtheorem{oracle}{Oracle}

\usepackage{color}
\usepackage{url}
\usepackage{amsthm}
\usepackage{times}

\usepackage{graphicx}
\usepackage{amssymb}
\usepackage{amsmath}
\usepackage{float}
\usepackage{enumerate}
\usepackage{algorithm,algorithmic}
\usepackage{multirow}

\newcommand{\tail}[2]{#1_{-#2}}
\newcommand{\Oh}{\mathcal{O}}

\newcommand{\loglog}{\mathrm{loglog}}
\newcommand{\polylog}{\mathrm{polylog}}

\newcommand{\abs}[1]{\left|#1\right|}

\newcommand{\norm}[2]{\left \lVert#2\right \rVert_{#1}}
\newcommand{\head}[2]{#1_{[#2]}}
\newcommand{\range}[3]{#1_{\overline{[#2]}\cap[#3]}}

\newcommand{\gaplinf}{\mathsf{Gap}\ell_{\infty}}

\newcommand{\indlinf}{\mathsf{Ind}\ell_{\infty}}

\DeclareMathOperator{\poly}{poly}
\def\R{\mathbb{R}}
\def\eps{\epsilon}

{\makeatletter
 \gdef\xxxmark{%
   \expandafter\ifx\csname @mpargs\endcsname\relax 
     \expandafter\ifx\csname @captype\endcsname\relax 
       \marginpar{xxx}
     \else
       xxx 
     \fi
   \else
     xxx 
   \fi}
 \gdef\xxx{\@ifnextchar[\xxx@lab\xxx@nolab}
 \long\gdef\xxx@lab[#1]#2{{\bf [\xxxmark #2 ---{\sc #1}]}}
 \long\gdef\xxx@nolab#1{{\bf [\xxxmark #1]}}
}

\begin{document}

\maketitle

\begin{abstract}
In the problem of adaptive compressed sensing, one wants to estimate an approximately $k$-sparse vector $x\in\mathbb{R}^n$ from $m$ linear measurements $A_1 x, A_2 x,\ldots, A_m x$, where $A_i$ can be chosen based on the outcomes $A_1 x,\ldots, A_{i-1} x$ of previous measurements. The goal is to output a vector $\hat{x}$ for which $$\|x-\hat{x}\|_p \le C \cdot \min_{k\text{-sparse } x'} \|x-x'\|_q,$$ with probability at least $2/3$,
where $C > 0$ is an approximation factor. Indyk, Price and Woodruff (FOCS'11) gave an algorithm for $p=q=2$ for $C = 1+\epsilon$
with $\Oh((k/\epsilon) \loglog (n/k))$ measurements and $\Oh(\log^*(k) \loglog (n))$ rounds
of adaptivity.
We first improve their bounds, obtaining a scheme with
$\Oh(k \cdot \loglog (n/k) + (k/\epsilon) \cdot \loglog(1/\epsilon))$ measurements and $\Oh(\log^*(k) \loglog (n))$ rounds,
as well as a scheme with $\Oh((k/\epsilon) \cdot \loglog (n\log (n/k)))$ measurements and an optimal $\Oh(\loglog (n))$ rounds.
We then provide novel adaptive 
compressed sensing
schemes with improved bounds for $(p,p)$ for every $0 < p < 2$. 
We show that the improvement from $O(k \log(n/k))$ measurements to $O(k \log \log (n/k))$ measurements 
in the adaptive setting can persist with a better $\epsilon$-dependence
for other values of $p$ and $q$. For example, when $(p,q) = (1,1)$, we obtain
$O(\frac{k}{\sqrt{\epsilon}} \cdot \log \log n \log^3 (\frac{1}{\epsilon}))$ measurements. 
We obtain
nearly matching lower bounds, showing our algorithms are close to optimal. Along the way, 
we also obtain the first nearly-optimal bounds for $(p,p)$ schemes for every $0 < p < 2$ even  
in the non-adaptive setting. 
\end{abstract}

\section{Introduction}
Compressed sensing, also known as sparse recovery, is a central object of study in data stream algorithms, with applications to monitoring network traffic~\cite{haupt2008compressed}, analysis of genetic data~\cite{saz09,kbgsw10}, and many other domains~\cite{m05}. The problem can be stated as recovering an underlying signal $x\in\R^n$ from \emph{measurements} $A_1 \cdot x,...,A_m\cdot x$ with the $C$-approximate $\ell_p/\ell_q$
recovery guarantee being 
\begin{equation}\label{eqn:guarantee}
\|x-\hat{x}\|_p\le C\min_{k\text{-sparse }x'}\|x-x'\|_q,
\end{equation}
where the $A_i$ are drawn from a distribution and $m\ll n$. 
The focus of this work is on {\it adaptive compressed sensing}, 
in which the measurements are chosen in rounds, and the choice of measurement
in each round depends on the outcome of the measurements in previous rounds.

Adaptive compressed sensing
has been studied in a number of different works \cite{JXC,CHNR,HCN,HBCN,MSW,AWZ,indyk2011power,price2013lower} in theoretical computer science, machine learning, image processing, and many other domains~\cite{indyk2011power,price2013lower,awasthi2016learning}. In theoretical computer science and machine learning, adaptive compressed sensing serves as an important tool to obtain sublinear algorithms for active learning in both time and space~\cite{indyk2011power,gilbert2012approximate,price2013lower,awasthi2016learning}. In image processing, the study of adaptive compressed sensing has led to compressed acquisition of sequential images with various applications in celestial navigation and attitude determination~\cite{gupta2012compressive}.

Despite a large amount of works on adaptive compressed sensing, the power of adaptivity remains a long-standing open problem.
Indyk, Price, and Woodruff \cite{indyk2011power} were the first to show that without any assumptions on the
signal $x$, one can obtain a number $m$ of measurements which is a $\log (n) / \log \log(n)$ factor smaller than what can be achieved
in the non-adaptive setting. Specifically, for $p = q = 2$ and $C = 1+\epsilon$, they show that
$m = \Oh(\frac{k}{\epsilon} \log \log (n))$ measurements suffice to achieve guarantee \eqref{eqn:guarantee}, whereas it is known
that any non-adaptive scheme requires $k = \Omega(\frac{k}{\epsilon} \log(\frac{n}{k}))$ measurements, provided
$\epsilon > \sqrt{\frac{k \log n}{n}}$ (Theorem 4.4 of \cite{price20111+}, see also \cite{DIPW}).
Improving the sample complexity as much as possible is desired, as it might correspond to, e.g., the amount of radiation a hospital patient is exposed to, or the amont of time a patient must be present for diagnosis. 

The $\ell_1/\ell_1$ problem was studied in \cite{price20111+}, for which perhaps surprisingly, a better dependence on 
$\epsilon$ was obtained than is possible for $\ell_2/\ell_2$ schemes. Still, the power of adaptivity for the $\ell_1/\ell_1$ recovery problem over its non-adaptive counterpart has remained unclear. An $O(\frac{k}{\sqrt{\epsilon}}\log n\log^3(\frac{1}{\epsilon}))$ non-adaptive bound was shown in \cite{price20111+}, while an adaptive lower bound of $\Omega(\frac{k}{\sqrt{\epsilon}}/\log\frac{k}{\sqrt{\epsilon}})$ was shown in \cite{price2013lower}. 
Recently several works \cite{sy16,mn16} have looked at other values of $p$ and $q$, even those for which $0 < p, q < 1$, which
do not correspond to normed spaces. The power of adaptivity for such error measures is also unknown. 

%

\subsection{Our Results}
Our work studies the problem of adaptive compressed sensing by providing affirmative answers to the above-mentioned open questions. We improve over the best known results for $p = q = 2$, and then provide novel adaptive compressed sensing guarantees for $0 < p = q < 2$ for every $p$ and $q$. See Table \ref{table:comparison of sample complexity} for a comparison of results.

\begin{table*}
\caption{The sample complexity of adaptive compressed sensing. Results without any citation given correspond to our new results.} \label{table:comparison of sample complexity}
\begin{center}
\begin{tabular}{p{1.8cm}||ll|p{2.8cm}}
\hline
$C$, Guarantees  & Upper Bounds  & Rounds  & Lower Bounds\\
\hline\hline
$1+\epsilon$, $\ell_1/\ell_1$ &
$\Oh(\frac{k}{\sqrt{\epsilon}} \loglog (n) \log^{\frac{5}{2}} (\frac{1}{\epsilon}))$ & $\Oh(\loglog (n))$ &
$\Omega(\frac{k}{\sqrt{\epsilon}\log (k/\sqrt{\epsilon}))})$~\cite{price2013lower}\\
\hline
$1+\epsilon$, $\ell_p/\ell_p$ & $\Oh(\frac{k}{\epsilon^{p/2}}\loglog(n) \poly(\log(\frac{1}{\eps})))$ & $\Oh(\loglog (n))$ & $\Omega(\frac{k}{{\eps}^{p/2}}\frac{1}{\log^2 (k/\eps)})$\\
\hline
$\sqrt{\frac{1}{k}}$, $\ell_{\infty}/\ell_2$ & $\Oh(k \loglog (n) + k \log (k) )$ & $\Oh(\loglog (n))$ &  -\\
\hline
\multirow{3}{2.5cm}{$1+\epsilon, \ell_2/\ell_2$} & $ \Oh(\frac{k}{\epsilon}\loglog (\frac{n\epsilon}{k}))$~\cite{indyk2011power} &  $\Oh(\log^*(k)\loglog(\frac{n\epsilon}{k}))$~\cite{indyk2011power} & \multirow{3}{3.2cm}{$\Omega(\frac{k}{\epsilon}+\loglog (n)$)~\cite{price2013lower}} \\
& $ \Oh(k \loglog (\frac{n}{k}) + \frac{k}{\epsilon} \loglog (\frac{1}{\epsilon}))$ & $\Oh(\log^*(k) \loglog (\frac{n}{k}))$ & \\
& $ \Oh(\frac{k}{\epsilon}\loglog(\frac{n\log (n\epsilon)}{k}))$ & $\Oh(\loglog (n\log (\frac{n \epsilon}{k}))$ & \\
\hline
\end{tabular}
\end{center}
\end{table*}

For $\ell_1/\ell_1$, we design an adaptive algorithm which requires only $\Oh(\frac{k}{\sqrt{\epsilon}}\loglog (n) \log^{\frac{5}{2}}(\frac{1}{\epsilon}))$ measurements for the $\ell_1/\ell_1$ problem. More generally, we study the $\ell_p/\ell_p$ problem for $0< p<2$. One of our main theorems is the following.

\begin{theorem}[$\ell_p/\ell_p$ Recovery Upper Bound]
\label{theorem: lplp recovery upper bound}
Let $x\in\R^n$ and $0 < p < 2$. There exists a randomized algorithm that performs $\Oh(\frac{k}{\epsilon^{p/2}}\loglog (n) \poly(\log (\frac{1}{\eps})))$ adaptive linear measurements on $x$ in $\Oh(\loglog (n))$ rounds, and with probability $2/3$, returns a vector $\hat{x} \in \R^n$ such that
$\norm{p}{x- \hat{x}} \leq (1+\eps) \|x_{-k}\|_p.$
\end{theorem}

Theorem \ref{theorem: lplp recovery upper bound} improves the previous sample complexity upper bound for the case of $C=1+\epsilon$ and $p=q=1$ from $\Oh(\frac{k}{\sqrt{\epsilon}}\log (n)\log^3(\frac{1}{\epsilon}))$ to $\Oh(\frac{k}{\sqrt{\epsilon}}\loglog (n)\log^{\frac{5}{2}}(\frac{1}{\epsilon}))$. Compared with the non-adaptive $(1+\epsilon)$-approximate $\ell_1/\ell_1$ upper bound of $\Oh(\frac{k}{\sqrt{\epsilon}}\log (n)\log^3(\frac{1}{\epsilon}))$, we show that adaptivity exponentially improves the sample complexity w.r.t. the dependence on $n$ over non-adaptive algorithms while retaining the improved dependence on $\epsilon$ of non-adaptive algorithms. Furthermore, Theorem \ref{theorem: lplp recovery upper bound} extends the working range of adaptive compressed sensing from $p=1$ to general values of $p\in(0,2)$.

We also state a complementary lower
bound to formalize the hardness of the above problem. 
\begin{theorem}[$\ell_p/\ell_p$ Recovery Lower Bound]
\label{theorem: lplp recovery lower bound}
Fix $0 < p < 2$, any $(1+\epsilon)$-approximate $\ell_p/\ell_p$ recovery scheme with
  sufficiently small constant failure probability must make
  $\Omega(\frac{k}{{\eps}^{p/2}}/\log^2 (\frac{k}{\eps}))$ measurements.
\end{theorem}
Theorem \ref{theorem: lplp recovery lower bound} shows that our upper bound in Theorem \ref{theorem: lplp recovery upper bound} is tight up to the $\log(k/\epsilon)$ factor. 

We also study the case when $p\neq q$. In particular, we focus on the case when $p=\infty,q=2$ and $C=\sqrt{\frac{1}{k}}$, as in the following theorem.

\begin{theorem}[$\ell_\infty/\ell_2$ Recovery Upper Bound]
\label{theorem: l_inftyl2 recovery lower bound}
Let $x \in \mathbb{R}^n$. There exists a randomized algorithm that performs $\Oh(k \log (k) + k \loglog (n))$ linear measurements on $x$ in $\Oh(\loglog (n))$ rounds, and with probability $1- 1/\mathrm{poly}(k)$ returns a vector $\hat{x}$ such that
$\|x-\hat{x}\|_{\infty}^2 \leq \frac{1}{k} \|x_{-k}\|_2^2$, 
where $x_{-k}\in\R^n$ is the vector with the largest $n-k$ coordinates (in the sense of absolute value) being zeroed out.
\end{theorem}
We also provide an improved result for $(1+\epsilon)$-approximate $\ell_2/\ell_2$ problems.

\begin{theorem}[$\ell_2/\ell_2$ Sparse Recovery Upper Bounds]
\label{theorem: l2l2 improved}
Let $x \in \mathbb{R}^n$. There exists a randomized algorithm that
\begin{itemize}
\item
uses $\Oh(\frac{k}{\epsilon} \loglog(\frac{1}{\epsilon}) + k \loglog (\frac{n}{k}))$ linear measurements on $x$ in $\Oh(\loglog (\frac{n}{k}) \cdot \log^*(k) )$ rounds;
\item
uses $\Oh(\frac{k}{\epsilon}\loglog(\frac{n\log (n \epsilon)}{k}))$ linear measurements on $x$ in $\Oh(\loglog (\epsilon n \log (\frac{n}{k})))$ rounds;
\end{itemize}
and with constant probability returns a vector $\hat{x}$ such that
$\|x-\hat{x}\|_2 \leq (1+\epsilon) \|x_{-k}\|_2$. 
\end{theorem}
Previously the best known tradeoff was $\Oh(\frac{k}{\epsilon}\loglog(\frac{n\epsilon}{k}))$ samples and $\Oh(\log^*(k) \loglog(\frac{n\epsilon}{k}))$ rounds for $(1+\epsilon)$-approximation for the $\ell_2/\ell_2$ problem~\cite{indyk2011power}. Our result improves both the sample complexity (the first result) and the number of rounds (the second result). We summarize our results in Table \ref{table:comparison of sample complexity}.

\subsection{Our Techniques}
\medskip
\noindent{\textbf{$\boldsymbol{\ell_\infty/\ell_2}$ Sparse Recovery.}}
Our $\ell_{\infty}/\ell_2$ sparse recovery scheme hashes every $i \in [n]$ to $\mathrm{poly}(k)$ buckets, and then proceeds by finding all the buckets that have $\ell_2$ mass at least $\Omega(\frac{1}{\sqrt{k}} \|x_{-\Omega(k)}\|_2)$. Clearly, there are $\Oh(k)$ of such buckets, and since all $k$ heavy coordinates are isolated due to hashing, we can find a set of buckets that contain all heavy coordinates, and moreover all these heavy coordinates are isolated from each other. Then, we run a $1$-sparse recovery in each bucket in parallel in order to find all the heavy coordinate. However, since we have $\Oh(k)$ buckets, we cannot afford to take a union bound over all one-sparse recovery routines called. Instead, we show that most buckets succeed and hence we can substract from $x$ the elements returned, and then run a standard \textsc{CountSketch} algorithm to recover everything else. This algorithm obtains an optimal $\Oh(\loglog (n))$ number of rounds and $\Oh(k \log (k) + k \loglog (n))$ number of measurements, while succeeding with probability at least $1- 1/\mathrm{poly}(k)$.

We proceed by showing an algorithm for $\ell_2/\ell_2$ sparse recovery with $\Oh( \frac{k}{\epsilon} \loglog (n))$ measurements and $\Oh(\loglog (n))$ rounds. This will be important for our more general $\ell_p/\ell_p$ scheme, saving a $\log^*(k)$ factor from the number of rounds, achieving optimality with respect to this quantity. For this scheme, we utilize the $\ell_{\infty}/\ell_2$ scheme we just developed, observing that for small $k < \Oh(\log (n))$, the measurement complexity is $\Oh(k \loglog (n))$. Our idea is then to exploit the fact that we can reduce the problem to smaller instances with logarithmic sparsity. The algorithm hashes to $k/(\epsilon \log (n))$ buckets, and in each bucket runs $\ell_{\infty}/\ell_1$ with sparsity $k/\epsilon$. Now, in each bucket there exist at most $\log (n)$ heavy elements, and the noise from non-heavy elements is ``low'' enough. The $\ell_{\infty}/\ell_2$ algorithm in each bucket succeeds with probability $1 - 1/\polylog (n))$; this fact allows us to argue that all but a $1/\polylog(n)$ fraction of the buckets will succeed, and hence we can recover all but a $k/\polylog (n))$ fraction of the heavy coordinates. The next step is to subtract these coordinates from our initial vector, and then run a standard $\ell_2/\ell_2$ algorithm with decreased sparsity.

\medskip
\noindent{\textbf{$\boldsymbol{\ell_p/\ell_p}$ Sparse Recovery.}}
Our $\ell_p/\ell_p$ scheme, $0< p < 2$, is based on carefully invoking
several $\ell_2/\ell_2$ schemes with different parameters.
We focus our discussion on $p = 1$, then
mention extensions to general $p$. A main difficulty of adapting
the $\ell_1/\ell_1$ scheme of \cite{price20111+} is that it relies upon an
$\ell_{\infty}/\ell_2$ scheme, and all known schemes, including ours,
have at least a $k \log k$ dependence on the number of measurements, which
is too large for our overall goal.

A key insight in \cite{price20111+} for $\ell_1/\ell_1$ is that since the output does not need to be exactly $k$-sparse, one can compensate for mistakes on approximating the top $k$ entries of $x$ by accurately outputting enough smaller entries. For example, if $k = 1$, consider two possible signals $x = (1, \epsilon, \ldots, \epsilon)$ and $x' = (1+\epsilon, \epsilon, \ldots, \epsilon)$, where $\epsilon$ occurs $1/\epsilon$ times in both $x$ and $x'$. One can show, using known lower bound techniques, that distinguishing $x$ from $x'$ requires $\Omega(1/\epsilon)$ measurements. Moreover, $x_1 = (1, 0, \ldots, 0)$ and $x'_1 = (1+\epsilon, 0, \ldots, 0)$, and any $1$-sparse approximation to $x$ or $x'$ must therefore distinguish $x$ from $x'$, and so requires $\Omega(1/\epsilon)$ measurements. An important insight though, is that if one does not require the output signal $y$ to be $1$-sparse, then one can output $(1, \epsilon, 0, \ldots, 0)$ in both cases, without actually distinguishing which case one is in!

As another example, suppose that $x = (1, \epsilon, \ldots, \epsilon)$ and $x' = (1+\epsilon^c, \epsilon, \ldots, \epsilon)$ for some $0 < c < 1$. In this case, one can show that one needs $\Omega(1/\epsilon^{c})$ measurements to distinguish $x$ and $x'$, and as before, to output an exactly $1$-sparse signal providing a $(1+\epsilon)$-approximation requires $\tilde{\Theta}(1/\epsilon^c)$ measurements. In this case if one outputs a signal $y$ with $y_1 = 1$, one cannot simply find a single other coordinate $\epsilon$ to ``make up'' for the poor approximation on the first coordinate. However, if one were to output $1/\epsilon^{1-c}$ coordinates each of value $\epsilon$, then the $\epsilon^c$ ``mass" lost by poorly approximating the first coordinate would be compensated for by outputting $\epsilon \cdot 1/\epsilon^{1-c} = \epsilon^c$ mass on these remaining coordinates. It is not clear how to find such remaining coordinates though, since they are much smaller; however, if one randomly subsamples an $\epsilon^c$ fraction of coordinates, then roughly $1/\epsilon^{1-c}$ of the coordinates of value $\epsilon$ survive and these could all be found with a number of measurements proportional to $1/\epsilon^{1-c}$. Balancing the two measurement complexities of $1/\epsilon^c$ and $1/\epsilon^{1-c}$ at $c = 1/2$ gives roughly the optimal $1/\epsilon^{1/2}$ dependence on $\epsilon$ in the number of measurements.

To extend this to the adaptive case, a recurring theme of the above examples
is that the top $k$, while they need to be found, they do not need to be
approximated very accurately. Indeed, they do need to be found, if, e.g.,
the top $k$ entries of $x$ were equal to an arbitrarily large value
and the remaining entries were much smaller. We accomplish this by running
an $\ell_2/\ell_2$ scheme with parameters $k' = \Theta(k)$ and
$\epsilon' = \Theta(\sqrt{\epsilon})$, as well as an $\ell_2/\ell_2$
scheme with parameters $k' = \Theta(k/\sqrt{\epsilon})$ and
$\epsilon' = \Theta(1)$ (up to logarithmic factors in $1/\epsilon$). Another
theme is that the mass in the smaller coordinates we find to compensate
for our poor approximation in the larger coordinates also does not need
to be approximated very well, and we find this mass by subsampling many
times and running an $\ell_2/\ell_2$ scheme with parameters $k' = \Theta(1)$
and $\epsilon' = \Theta(1)$. This technique is surprisingly general, 
and does not require the underlying error measure we are approximating
to be a norm. It just uses scale-invariance and how its rate of growth
compares to that of the $\ell_2$-norm. 

\medskip
\noindent{\textbf{$\boldsymbol{\ell_2/\ell_2}$ Sparse Recovery.}}
Our last algorithm, which concerns $\ell_2/\ell_2$ sparse recovery, achieves $\Oh(k \loglog (n) + \frac{k}{\epsilon} \loglog (1/\epsilon))$ measurements, showing that $\epsilon$ does not need to multiply $\loglog (n)$. The key insight lies in first solving the $1$-sparse recovery task with $\Oh(\loglog (n) + \frac{1}{\epsilon} \loglog (1/\epsilon))$ measurements, and then extending this to the general case. To achieve this, we hash to $\polylog (1/\epsilon)$ buckets, then solve $\ell_2/\ell_2$ with constant sparsity on a new vector, where coordinate $j$ equals the $\ell_2$ norm of the $j$th bucket; this steps requires only $\Oh(\frac{1}{\epsilon} \loglog (1/\epsilon))$ measurements. Now, we can run standard $1$-sparse recovery in each of these buckets returned. Extending this idea to the general case follows by plugging this sub-routine in the iterative algorithm of \cite{indyk2011power}, while ensuring that sub-sampling does not increase the number of measurements. This means that we have to sub-sample at a slower rate, slower roughly by a factor of $\epsilon$. The guarantee from our $1$-sparse recovery algorithm fortunately allows this slower sub-sampling to go through and give the desired result.

\medskip
\noindent{\textbf{Notation:}} For a vector $x \in \mathbb{R}^n$, we define $H_k(x)$ to be the set of its largest $k$ coordinates in absolute value. For a set $S$, denote by $x_S$ the vector with every coordinate $i \notin S$ being zeroed out. We also define $x_{-k} = x_{[n] \setminus H_k(x)}$ and $ H_{k,\epsilon}(x) = \{ i \in [n]: |x_i| \geq \frac{\epsilon}{k} \|x_{-k}\|_2^2\}$, where $[n]$ represents the set $\{1,2,...,n\}$. For a set $S$, let $|S|$ be the cardinality of $S$.

Due to space constraints, we defer the proof of Theorem \ref{theorem: lplp recovery lower bound} to the appendix. 

\section{Adaptive $\ell_p/\ell_p$ Recovery}
This section is devoted to proving Theorem \ref{theorem: lplp recovery upper bound}. Our algorithm for $\ell_p/\ell_p$ recovery is in Algorithm \ref{algorithm: the sketching algorithm for even p}.

Let $f = {\eps}^{p/2}$, ${r= 2/(p\log (1/f))}$ and $q = \max\{p-\frac{1}{2},0\}= (p-\frac{1}{2})^+$. We will invoke the following $\ell_2/\ell_2$ oracle frequently throughout the paper.
\begin{oracle}[$\textsc{AdaptiveSparseRecovery}_{\ell_p/\ell_q}(x,k,\epsilon)$]
\label{oracle}
The oracle is fed with $(x,k,\epsilon)$ as input parameters, and outputs a set of coordinates $i\in[n]$ of size $\mathcal{O}(k)$ which corresponds to the support of vector $\hat{x}$, where $\hat{x}$ can be any vector for which $\|x-\hat{x}\|_p\le (1+\epsilon)\min_{\mathcal{O}(k)\text{-sparse }x'}\|x-x'\|_q$.
\end{oracle}

Existing algorithms can be applied to construct Oracle \ref{oracle} for the $\ell_2/\ell_2$ case, such as \cite{indyk2011power}. Without loss of generality, we assume that the coordinates of $x$ are ranked in decreasing value, i.e., $x_1\geq x_2\geq\cdots \geq x_n$.

\begin{algorithm}
\caption{Adaptive $\ell_p/\ell_p$ Recovery}
\label{algorithm: the sketching algorithm for even p}
\begin{algorithmic}
\STATE {\bfseries 1.} $A \leftarrow \textsc{AdaptiveSparseRecovery}_{\ell_2/\ell_2}(x,2k/f, 1/10).$
\STATE {\bfseries 2.} $B \leftarrow \textsc{AdaptiveSparseRecovery}_{\ell_2/\ell_2}(x,4k, f/r^2).$
\STATE {\bfseries 3.} $S \leftarrow A \cup B$.
\STATE {\bfseries 4.} \textbf{For} $j=1:r$
\STATE {\bfseries 5.} \quad Uniformly sample the entries of $x$ with probability $2^{-j}f/k$ for ${k/(2f(r+1)^q)}$ times.
\STATE {\bfseries 6.} \quad Run the adaptive $\textsc{AdaptiveSparseRecovery}_{\ell_2/\ell_2}(x,2,{1/(4(r+1))^{\frac{2}{p}}})$ algorithm on each of the ${k/(2f(r+1)^q)}$ subsamples to obtain sets $A_{j,1}, A_{j,2},\ldots, A_{j,{k/(2f(r+1)^q)}}$.
\STATE {\bfseries 7.} \quad Let $S_j \leftarrow \cup_{t=1}^{k/(2f(r+1)^{q})} A_{j,t} \setminus \cup_{t=0}^{j-1} S_t$.
\STATE {\bfseries 8.} \textbf{End For}
\STATE {\bfseries 9.} Request the entries of $x$ with coordinates $S_0,...,S_{r}$.
\STATE {\bfseries Output:} $\hat x=x_{S_0\cup\dotsb\cup S_r}$.
\end{algorithmic}
\end{algorithm}


\begin{lemma}\label{lemma:subsample 2-norm}
Suppose we subsample $x$ with probability $p$ and let $y$ be the subsampled vector formed from $x$. Then with failure probability $e^{-\Omega(k)}$,
$
\norm{2}{\tail{y}{2k}} \leq \sqrt{2p}\norm{2}{\tail{x}{k/p}}.
$
\end{lemma}

\begin{proof}
Let $T$ be the set of coordinates in the subsample. Then $\mathbb{E}\left[\left|T \cap {\left[\frac{3k}{2p}\right]}\right|\right] = \frac{3k}{2}$. So by the Chernoff bound,
$
\text{Pr} \left[ \left|T \cap {\left[\frac{3k}{2p}\right]}\right| > 2k \right] \leq e^{-\Omega{(k)}}.
$ Thus $ \left|T \cap {\left[\frac{3k}{2p}\right]}\right| \leq  2k$ holds with high probability.
Let $Y_i= x_i^2 $ if $i \in T$ $Y_i=0$ if $i\in [n]\setminus T$. Then
$\mathbb{E}\left[\sum_{i > \frac{3k}{2p}} Y_i\right] = p \norm{2}{\tail{x}{\frac{3k}{2p}}}^2 \leq p\norm{2}{\tail{x}{k/p}}^2.$ 
Notice that there are at least $\frac{k}{2p}$ elements in $\tail{x}{k/p}$ with absolute value larger than $\abs{x_{\frac{3k}{2p}}}$.
Thus for $i>\frac{3k}{2p}$,
$
Y_i \leq \abs{x_{\frac{3k}{2p}}}^2 \leq \frac{2p}{k} \norm{2}{\tail{x}{k/p}}^2.
$
Again by a Chernoff bound,
$
\text{Pr} \left[\sum_{i>\frac{3k}{2p}} Y_i \geq \frac{4p}{3} \norm{2}{\tail{x}{k/p}}^2\right] \leq e^{-\Omega(k)}.
$
Conditioned on the latter event not happening,
$
\norm{2}{\tail{y}{2k}}^2 \leq \sum_{i>\frac{3k}{2p}} Y_i \leq  \frac{4p}{3} \norm{2}{\tail{x}{k/p}}^2 \leq 2p\norm{2}{\tail{x}{k/p}}^2.
$
By a union bound, with failure probability $e^{-\Omega(k)}$, we have
$
\norm{2}{\tail{y}{2k}} \leq \sqrt{2p}\norm{2}{\tail{x}{k/p}}.
$
\end{proof}

\begin{lemma}\label{lemma:l2controlp-norm}
Let $\hat{x}$ be the output of the $\ell_2/\ell_2$ scheme on $x$ with parameters $(k, \eps/2)$. Then with small constant failure probability,
$\norm{p}{\head{x}{k}}^p - \norm{p}{\hat{x}}^p \leq k^{1-\frac{p}{2}}\eps^{\frac{p}{2}}\norm{2}{\tail{x}{k}}^p.$
\end{lemma}
\begin{proof}
Notice that with small constant failure probability, the $\ell_2/\ell_2$ guarantee holds and we have
\begin{equation*}
\begin{split}
\norm{2}{\head{x}{k}}^2 - \norm{2}{\hat{x}}^2
= \norm{2}{x - \hat{x}}^2-\norm{2}{\tail{x}{k}}^2
\leq
(1+\eps)\norm{2}{\tail{x}{k}}^2 - \norm{2}{\tail{x}{k}}^2   
= \eps\norm{2}{\tail{x}{k}}^2 .
\end{split}
\end{equation*}
Let $S\subset[n]$ be such that $x_S = \hat{x}$, and define $y = x_{[k]\setminus S}$, $z = x_{S\setminus [k]}$. Then if $\norm{p}{y}^p \leq k^{1-\frac{p}{2}}\eps^{\frac{p}{2}}\norm{2}{\tail{x}{k}}^p$ we are done. Otherwise, let $1\leq k'\leq k$ denote the size of $[k]\setminus S$, and define $c = \norm{2}{y}/{\sqrt{k'}}$. 
\begin{equation*}
\begin{split}
\norm{p}{\head{x}{k}}^p - \norm{p}{\hat{x}}^p
&= \norm{p}{y}^p - \norm{p}{z}^p
\leq {k'}^{1 - \frac{p}{2}}\norm{2}{y}^p - \norm{p}{z}^p
= \frac{\norm{2}{y}^2}{c^{2-p}} - \norm{p}{z}^p\\ &
\leq \frac{\norm{2}{y}^2 -  \norm{2}{z}^2}{c^{2-p}}
= \frac {\norm{2}{\head{x}{k}}^2 - \norm{2}{\hat{x}}^2}{c^{2-p}}\leq \frac{\eps\norm{2}{\tail{x}{k}}^2}{c^{2-p}}.
\end{split}
\end{equation*}
Since
$
c \geq \frac{\norm{p}{y}}{k'^{\frac{1}{p}}} \geq \frac{\norm{p}{y}}{k^{\frac{1}{p}}} \geq \sqrt{\frac{\eps}{k}}\norm{2}{\tail{x}{k}},
$
we have
$
\norm{p}{\head{x}{k}}^p - \norm{p}{\hat{x}}^p  \leq k^{\frac{2-p}{2}} \eps^{1-\frac{2-p}{2}} \norm{2}{\tail{x}{k}}^{2-(2-p)} = k^{1-\frac{p}{2}}\eps^{\frac{p}{2}}\norm{2}{\tail{x}{k}}^p.
$
\end{proof}

\begin{theorem}
Fix $0<p<2$. For $x\in\R^n$,
there exists a $(1+\eps)$-approximation algorithm that performs $\Oh(\frac{k}{\epsilon^{p/2}}\loglog (n) \log^{\frac{2}{p}+1-(p-\frac{1}{2})^+} (\frac{1}{\eps}))$ adaptive linear measurements in {$\Oh(\loglog (n))$} rounds, and with probability at least $2/3$, we can find a vector $\hat{x} \in \R^n$ such that
\begin{equation}
\norm{p}{x- \hat{x}} \leq (1+\eps) \norm{p}{\tail{x}{k}}.
\end{equation}
\end{theorem}

\begin{proof}
The algorithm is stated in Algorithm \ref{algorithm: the sketching algorithm for even p}.
We first consider the difference $\norm{p}{\head{x}{k}}^p - \norm{p}{x_{S_0}}^p$. \\
Let $i^*(0)$ be the smallest integer such that for any $l>i^*(0)$, $|x_{l}| \leq {\|x_{-{2k/f}}\|_2/\sqrt{k}}$.

Case 1. $i^*(0) > 4k$ \\
Then for all $k < j \leq 4k$, we have $|x_j| > \|x_{-2k/f}\|_2/ \sqrt{k}$.
Hence $x_{S_0}$ must contain at least $1/2$ of these indices; if not, the total squared loss is at least $1/2 \cdot 3k\|x_{-2k/f}\|_2^2/k \geq (3/2)\|x_{-2k/f}\|_2^2$, a contradiction to $\eps' =1/10$. It follows that
$
\|x_{S_0 \cap \{k+1, ..., 4k\}}\|_p^p \geq
\frac{3}{2} k \left[\frac{\|x_{-2k/f}\|_2}{\sqrt{k}}\right]^p
= \frac{3}{2} k^{1-\frac{p}{2}} \|x_{-2k/f}\|_2^p.
$
On the other hand, $\norm{p}{\head{x}{k}}^p - \norm{p}{x_{S_0}}^p$ is at most $1.1k^{1-\frac{p}{2}}\|x_{-{2k/f}}\|_2^p$,
since by the $\ell_2/\ell_2$ guarantee
$$
\|x_{[k]}\|_p^p - \|x_{S_0 \cap{[k]}}\|_p^p \leq k^{1-\frac{p}{2}}\|x_{[k]}-x_{S_0 \cap{[k]}}\|_2^p
\leq k^{1-\frac{p}{2}} \|x - x_{S_0}\|_2^p \leq \frac{11}{10} k^{1-\frac{p}{2}}\|x_{-{2k/f}}\|_2^p.
$$
It follows that
$\|x_{[k]}\|_p^p - \|x_{S_0}\|_p^p
=\|x_{[k]}\|_p^p - \|x_{S_0\cap {[k]}}\|_p^p - \|x_{S_0 \cap \{k+1, ..., 4k\}}\|_p^p 
\leq \frac{11}{10}k^{1-\frac{p}{2}} \|x_{-2k/f}\|_2^p - \frac{3}{2} k^{1-\frac{p}{2}} \|x_{-2k/f}\|_2^p 
\leq 0.$

Case 2. $i^*(0) \leq 4k$, and $\sum_{j=i^*(0)+1}^{2k/f} x_j^2\geq 4\|x_{-2k/f}\|_2^2$. \\
We claim that $x_{S_0}$ must contain at least a $5/8$ fraction of coordinates in $\{i^*(0)+1,..., 2k/f\}$; if not, then the cost for missing at least a $3/8$ fraction of the $\ell_2$-norm of $x_{\{i^*(0)+1,..., 2k/f\}}$ will be at least $(3/2)\|x_{-2k/f}\|_2^2$, contradicting the $\ell_2/\ell_2$ guarantee.
Since all coordinates $x_j$'s for $j > i^*(0)$ have value at most $\|x_{-2k/f}\|_2/\sqrt{k}$, it follows that the $p$-norm of coordinates corresponding to $\{i^*(0)+1,..., 2k/f\}\cap S_0$ is at least
$\norm{p}{x_{\{i^*(0)+1,..., 2k/f\}\cap S_0}}^p \geq \frac{5}{2} k^{\frac{2-p}{2}}\frac{\|x_{-2k/f}\|_2^2}{\|x_{-2k/f}\|_2^{2-p}} = \frac{5}{2}k^{1-\frac{p}{2}} \|x_{-2k/f}\|_2^p .$
Then
\begin{equation*}
\begin{split}
\|x_{[k]}\|_p^p - \|x_{S_0}\|_p^p &\leq \frac{11}{10}k^{1-\frac{p}{2}} \|x_{-2k/f}\|_2^p  + k\left(\frac{\|x_{-2k/f}\|_2}{\sqrt{k}}\right)^p -
\|x_{\{i^*(0)+1,..., 2k/f\}\cap S_0}\|_p^p\\
&\leq \frac{21}{10}k^{1-\frac{p}{2}} \|x_{-2k/f}\|_2^p  - \frac{5}{2}k^{1-\frac{p}{2}} \|x_{-2k/f}\|_2^p
\leq 0.
\end{split}
\end{equation*}

Case 3. $i^*(0) \leq 4k$, and $\sum_{j=i^*(0)+1}^{2k/f} x_j^2\leq 4\|x_{-2k/f}\|_2^2$. \\
With a little abuse of notation, let $x_{S_0}$ denote the output of the $\ell_2/\ell_2$ with parameters $(4k, f/r^2)$. Notice that there are at most $8k$ non-zero elements in $x_{S_0}$, and
$\|x_{-4k}\|_2^2 \leq \|x_{-i^*(0)}\|_2^2 = \sum_{j=i^*(0)+1}^{2k/f} x_j^2 + \|x_{-2k/f}\|_2^2 \leq 5\|x_{-2k/f}\|_2^2.$
By Lemma~\ref{lemma:l2controlp-norm},
we have
$\norm{p}{\head{x}{k}}^p - \norm{p}{x_{S_0}}^p \leq \norm{p}{\head{x}{4k}}^p - \norm{p}{x_{S_0}}^p
\leq (4k)^{1-\frac{p}{2}} \frac{f^{\frac{p}{2}}}{r^p}\|x_{-4k}\|_2^p \leq \Oh\left(\frac{1}{r^p}\right)k^{1-\frac{p}{2}} f^{\frac{p}{2}} \|x_{-2k/f}\|_2^p.$
According to the above three cases, we conclude that
$\|x_{[k]}\|_p^p - \|x_{S_0}\|_p^p \leq \Oh\left(\frac{1}{r^p}\right)k^{1-\frac{p}{2}} f^{\frac{p}{2}} \|x_{-2k/f}\|_2^p.$
Thus with failure probability at most $1/6$,
\begin{equation}
\|x - \hat{x}\|_p^p - \| x_{-k}\|_p^p = \|x_{[k]}\|_p^p - \sum_{j=0}^{r} \|x_{S_j}\|_p^p \leq \Oh\left(\frac{1}{r^p}\right)k^{1-\frac{p}{2}} f^{\frac{p}{2}} \|x_{-2k/f}\|_2^p- \sum_{j=1}^r \norm{p}{x_{S_j}}^p.\label{eq:totalerror}
\end{equation}
In order to convert the first term on the right hand side of \eqref{eq:totalerror} to a term related to the $\ell_p$
norm (which is a semi-norm if $0 < p < 1$), 
we need the following inequalities: for every $u$ and $s$, by splitting into chunks of size
$s$, we have
\begin{align*}
  s^{1 - \frac{p}{2}}\norm{2}{\tail{u}{2s}}^p \leq \norm{p}{\tail{u}{s}}^p,\qquad\mbox{and}\qquad
  \norm{2}{\range{u}{s}{2s}} \leq \sqrt{s}\abs{u_{s}}.
\end{align*}
Define $c = (r+1)^{\min\{p, 1\}}$. This gives us that, for $0< p<2$
%
$\frac{1}{(r+1)^p}k^{1-\frac{p}{2}} f^{\frac{p}{2}}\norm{2}{\tail{x}{2k/f}}^p \leq \frac{k^{1-\frac{p}{2}} f^{\frac{p}{2}}}{c}\norm{2}{\tail{x}{2k/f^{1+\frac{2}{p}}}}^p
+  \frac{k^{1-\frac{p}{2}} f^{\frac{p}{2}}}{c}\sum_{j=1}^{r} \norm{2}{\range{x}{2^jk/f}{2^{j+1}k/f}}^p
 \leq \frac{f^{(1-\frac{p}{2})(1+\frac{2}{p})+\frac{p}{2}}}{c}\norm{p}{\tail{x}{k/f^{1+\frac{2}{p}}}}^p + \frac{1}{c}\sum_{j=1}^{r} k2^{pj/2}\abs{x_{2^jk/f}}^p.$
Therefore,
\begin{align}
\label{eq:l2l1telescope}
\|\hat{x}-x\|_p^p-\|x_{-k}\|_p^p&\le \Oh\left(\frac{1}{c}\right)f^{\frac{2}{p}}\left\|x_{-k/f^{1+\frac{2}{p}}}\right\|_p^p+\sum_{j=1}^r \Oh\left(\frac{1}{c}\right)k2^{pj/2}|x_{2^jk/f}|^p-\sum_{j=1}^r \|x_{S_j}\|_p^p \notag\\
&\leq \Oh\left(\frac{1}{c}\right)f^{\frac{2}{p}}\left\|x_{-k/f}\right\|_p^p+\sum_{j=1}^r \Oh\left(\frac{1}{c}\right)k2^{pj/2}|x_{2^jk/f}|^p-\sum_{j=1}^r \|x_{S_j}\|_p^p .
\end{align}
Let $y=x_T$ denote an independent subsample of $x$ with probability $f/(2^j k)$, and $\hat{y}$ be the output of the $\ell_2/\ell_2$  algorithm with parameter s$(2, 1/(4(r+1))^{\frac{2}{p}})$. Notice that $|S_j| \leq 2k/(r+1)f$ by the adaptive $\ell_2/\ell_2$ guarantee. Define $Q = [2^{j}k/f] \setminus (S_0\cup \cdots \cup S_{j-1})$. There are at least $2^{j}k/(2f)$ elements in $Q$, and every element in $Q$ has absolute value at least $\abs{x_{2^jk/f}}$.
In each subsample, notice that
$\mathbb{E}[|T\cap Q|] = \frac{1}{2}$. Thus with sufficiently small constant failure probability there exists at least
$1$ element in $y$ with absolute value at least $|x_{2^{j}k/f}|$.
On the other hand, by Lemma~\ref{lemma:l2controlp-norm} and Lemma~\ref{lemma:subsample 2-norm},
\begin{equation}\label{controllp}
\begin{split}
\norm{p}{\head{y}{1}}^p - \norm{p}{\hat{y}}^p
\leq\norm{p}{\head{y}{2}}^p - \norm{p}{\hat{y}}^p
\leq \frac{2^{1-\frac{p}{2}}}{4(r+1)} \norm{2}{\tail{y}{2}}^p 
\leq \frac{1}{2(r+1)} \left(\frac{f}{2^jk}\right)^{\frac{p}{2}}\norm{2}{\tail{x}{2^j k /f}}^p,
\end{split}
\end{equation}
with sufficiently small constant failure probability given by the union bound. For the $k/(2f{(r+1)^{q}})$ independent copies of subsamples, by a Chernoff bound, a $1/4$ fraction of them will have the largest absolute value in $Q$ and \eqref{controllp} will also hold, with the overall failure probability being $e^{-\Omega{(k/(fr^q))}}$.
Therefore, since $k/f > 2^{pj/2} k$,
$\norm{p}{x_{S_j}}^p \geq \frac{2^{pj/2} k}{8(r+1)^{q}}\left[\abs{x_{2^j k/f}}^p - \frac{1}{2(r+1)} \left(\frac{f}{2^jk}\right)^{\frac{p}{2}}\norm{2}{\tail{x}{2^j k /f}}^p\right] 
\geq  \frac{2^{pj/2} k}{8(r+1)^{q}}\abs{x_{2^j k/f}}^p - \frac{k^{1-\frac{p}{2}}f^{\frac{p}{2}}}{16(r+1)^{q+1}} \norm{2}{\tail{x}{2k /f}}^p,$
and by the fact that $0<q<p<2$,
\begin{equation*}
\begin{split}
\|x - \hat{x}\|_p^p &- \| x_{-k}\|_p^p \leq
\Oh(\frac{1}{r^p})k^{1-\frac{p}{2}} f^{\frac{p}{2}} \|x_{-2k/f}\|_2^p- \sum_{j=1}^r \norm{p}{x_{S_j}}^p \\
&\leq \left[ \Oh\left(\frac{1}{r^p}\right) + \frac{r}{16(r+1)^{q +1}}\right]k^{1-\frac{p}{2}} f^{\frac{p}{2}} \|x_{-2k/f}\|_2^p - \sum_{j=1}^r  \frac{2^{pj/2} k}{8(r+1)^{q}}\abs{x_{2^j k/f}}^p\\
&\leq \Oh\left(\frac{1}{c}\right) f^{\frac{2}{p}}\norm{p}{\tail{x}{k/f}}^p +\left[ \Oh\left(\frac{1}{c}\right)+ \frac{1}{16(r+1)^{q}}-\frac{1}{8(r+1)^{q}}\right]\sum_{j=1}^{r} k2^{pj/2}\abs{x_{2^jk/f}}^p\\
&\leq f^{\frac{2}{p}}\norm{p}{\tail{x}{k/f}}^p \leq \eps \norm{p}{\tail{x}{k}}^p.
\end{split}
\end{equation*}
The total number of measurements will be at most
$$
\Oh\left(\frac{k}{f} \loglog (n)\hspace{-0.1cm} + \hspace{-0.1cm}\frac{4kr^2}{f} \loglog (n)\hspace{-0.1cm}+\hspace{-0.1cm} \frac{kr}{2fr^{q}} r^{\frac{2}{p}} \loglog (n)\right) = \Oh\left(\frac{k}{\epsilon^{\frac{p}{2}}}\loglog (n) \log^{\frac{2}{p}+1-(p-\frac{1}{2})^+}\hspace{-0.1cm} \left(\frac{1}{\eps}\right)\right),
$$
while the total failure probability given by the union bound is $1/6 +e^{-\Omega{(k/(fr^q))}}<1/3$,
which completes the proof.
\end{proof}

\section{$\ell_\infty/\ell_2$ Adaptive Sparse Recovery}

In this section, we will prove Theorem \ref{theorem: l_inftyl2 recovery lower bound}.
Our algorithm first approximates $\|x_{-k}\|_2$. The goal is to compute a value $V$ which is not much smaller than $\frac{1}{k} \|x_{-k}\|_2^2$, and also at least $\Omega(\frac{1}{k}) \|x_{-\Omega(k)}\|_2^2 $. This value will be used to filter out coordinates that are not large enough, while ensuring that heavy coordinates are included. We need the following lemma, which for example can be found in Section 4 of \cite{li2017sublinear}.

\begin{lemma}
Using $\log(1/\delta)$ non-adaptive measurements we can find with probability $1-\delta$ a value $V$ such that $ \frac{1}{C_1k} \|x_{-C_2k}\|_2^2 \leq V \leq \frac{1}{k}\|x_{-k}\|_2^2,		$ where $C_1,C_2$ are absolute constants larger than $1$.
\end{lemma}

We use the aforementioned lemma with $\Theta(\log k)$ measuremenents to obtain such a value $V$ with probability $1-1/\mathrm{poly}(k)$. Now let $c$ be an absolute constant and let $ g:[n] \rightarrow [k^c]$ be a random hash function. Then, with probability at least $1 - \frac{1}{\mathrm{poly}(k)}$ we have that for every $i,j \in H_k(x)$, $g(i) \neq g(j)$. By running \textsc{PartitionCountSketch}$(x,2C_1k,\{g^{-1}(1), g^{-1}(2),\ldots g^{-1}(k^c)\}$, we get back an estimate $w_j$ for every $j \in [k^c]$; here $C_1$ is an absolute constant. Let $\gamma'$ be an absolute constant to be chosen later. We set $ S = \{j \in [k^c]: w_j^2 \geq \gamma' V\} $ and $T = \bigcup_{j \in S} g^{-1}(j).$ We prove the following lemma.

\begin{lemma}

Let $C'$ be an absolute constant. With probability at least $1 - 1/\mathrm{poly}(k)$ the following holds.
\begin{enumerate}
\item  $|S| = \Oh(k)$.
\item Every $j \in [k^c]$ such that there exists $i \in H_k(x) \cap g^{-1}(j)$, will be present in $S$.
\item For every $j \in S$, there exists exactly one coordinate $i \in g^{-1}(j)$ with $x_i^2 \geq \frac{1}{C'k} \|x_{-C_2k}\|_2^2$.
\item  For every $j \in S$, $\|x_{ g^{-1}(j) \setminus H_k(x) }\|_2^2 \leq \frac{1}{k^{2}} \|x_{-k}\|_2^2$.
\end{enumerate}
\end{lemma}

\begin{proof}

Let $C_0$ be an absolute constant larger than $1$. Note that with probability $1 - C_0^2\cdot k^{6-c}$, all $i \in H_{C_0k^3}(x)$ (and, hence, also in $H_{C_0 k^3,1/k^3}(x)$) are isolated under $g$. Fix $j \in [k^c]$ and, for $i\in[n]$, define the random variable $Y_i = 1_{g(x_i) = j} x_i^2$. Now observe that

		\[\mathbb{E}\left [ \sum_{i \in g^{-1}(j) \setminus H_{C_0k^3,1/k^3}(x)} Y_i \right]= \frac{1}{k^c} \|x_{-C_0 k^3}\|_2^2. \]
Applying Bernstein's inequality to the variables $Y_i$ with
	\[	K = \frac{1}{C_0 k^3}\|x_{-C_0k^3}\|_2^2,\qquad\text{and}\qquad \sigma^2 < \frac{1}{k^{c+3}}\|x_{-C_0k^3}\|_2^4,\]
 we have that \[	\Pr \left[ \sum_{i \in g^{-1}(j) \setminus H_{C_0k^3,1/k^3}(x) }x_i^2 \geq 1/k^2 \|x_{-C_0k^2}\|_2^2 \right] \leq e^{-k}, \]
where $c$ is an absolute constant. This allows us to conclude that the above statement holds for all different $k^c$ possible values $j$, by a union-bound. We now prove the bullets one by one. We remind the reader that \textsc{PartitionCountSketch} aproximates the value of every $\|x_{g^{-1}(j)}\|_2^2$ with a multiplicate error in $[1-\gamma,1+\gamma]$ and additive error $\frac{1}{C_0k}\|x_{-k}\|_2^2$.\newline

1. Since there are at most $\frac{1}{\gamma'(1+\gamma)}C_2k + C_2k$ indices $j$ with $(1+\gamma)\|x_{g^{-1}(j)}\|_2^2 \geq \frac{\gamma'}{k} \|x_{-k}\|_2^2 \geq \gamma' V$, the algorithm can output at most $\Oh(k)$ indices.\newline

2. The estimate for such a $j$ will be at least $(1-\gamma)\frac{1}{k}\|x_{-k}\|_2^2 - \frac{1}{2C_1k} \|x_{-C_2k}\|_2^2 \geq \gamma' V$, for some suitable choice of $\gamma'$. This implies that $j$ will be included in $S$.\newline

3. Because of the guarantee for $V$ and the guarantee of \textsc{PartitionCountSketch}, we have that all $j$ that are in $S$ satisfy $(1+\gamma)\|x_{g^{-1}(j)} \|_2^2 + \frac{1}{k}\|x_{-2C_1k}\|_2^2 \geq \frac{\gamma'}{k}\|x_{-C_2k}\|_2^2$, and since
\[\sum_{i \in g^{-1}(j) \setminus H_{C_0k^3}(x)} x_i^2 \leq \frac{1}{k^2} \|x_{-k}\|_2^2, \] this implies that there exists $i \in H_{C_0k^3}(x) \cap g^{-1}(j)$. But since all $i \in H_{C_0k^3}(x)$ are perfectly hashed under $g$, this implies that this $i$ should satisfy $x_i^2 \geq \frac{1}{C'k}\|x_{-C_2k}\|_2^2$, from which the claim follows.\newline

4. Because elements in $H_{C_0k^3}(x)$ are perfectly hashed, we have that

	\[ \|x_{g^{-1}(j) \setminus H_k(x)}\|_2^2 = \|x_{g^{-1}(j) \setminus H_{C_0k^3}}(x) \|_2^2 \leq \frac{1}{k^2} \|x_{-k}\|_2^2 \]
for $C_0$ large enough.
\end{proof}

Given $S$, we proceed in the following way. For every $j \in S$, we run the algorithm from Lemma \ref{onesparse} to obtain an index $i_j$, using $\Oh(k \loglog n)$ measurements. Then we observe directly $x_{i_j}$  using another $\Oh(k)$ measurements, and form vector $z = x - x_{ \{ i_j \}_{j \in S} }$. We need the following lemma.

\begin{lemma}
With probability $1- 1/\mathrm{poly}(k)$, $|H_k(x) \setminus \{i_j\}_{j \in S} | \leq \frac{k}{\log ^2n }$.
\end{lemma}

\begin{proof}
Let us consider the calls to the 1-sparse recovery routine in $j$ for which there exists $i \in H_k(x) \cap g^{-1}(j)$. Since the 1-sparse recovery routine succeeds with probability $1-1/\mathrm{poly}(\log n)$, then the probability that we have more than $\frac{k}{\log^2n}$ calls that fail, is
	\[	{ k \choose \frac{k}{\log^2n}  } \left(\frac{1}{\mathrm{poly}(\log n)}\right)^{k/\log^2 n} \leq \frac{1}{\mathrm{poly}(k)}.	\]
This gives the proof of the lemma.
\end{proof}

For the last step of our algorithm, we run 
\textsc{PartitionCountSketch}$(z_T,k/ \log (n), [n])$ to estimate the entries of $z$. We then find the coordinates with the largest $2k$ estimates, and observe them directly. Since
\[ \frac{\log n}{k}\|(z_T)_{-k/\log n}\|_2^2 \leq \frac{\log n}{k} \cdot \frac{1}{k^2}\|x_{-k}\|_2^2 = \frac{\log n}{k^3} \|x_{-k}\|_2^2, \]
every coordinate will be estimated up to additive error $\frac{\log n}{k^3}\|x_{-k}\|_2^2$, which shows that every coordinate in $T\cap H_{k,1/k}(x)$ will be included in the top $2k$ coordinates. Putting everything together, it is guaranteed that we have found every coordinate $i \in S$. Moreover, since all lemmas hold with probability $1-1/\mathrm{poly}(k)$, the failure probability is $1/\mathrm{poly}(k)$ and the number of rounds is $\Oh( \loglog n)$.

\section{$\ell_2/\ell_2$ Adaptive Sparse Recovery in Optimal Rounds}

In this section, we give an algorithm for $\ell_2/\ell_2$ compressed sensing using $\Oh(\loglog n)$ rounds, instead of $\Oh(\log^*k \cdot \loglog n)$ rounds. Specifically, we prove the first bullet of Theorem~\ref{theorem: l2l2 improved}. We call this algorithm $\textsc{AdaptiveSparseRecovery}_{\ell_{\infty}/\ell_2}$.


We proceed with the design and the analysis of the algorithm. We note that for $k/\epsilon = \Oh(\log^5 n)$\footnote{the constant $5$ is arbitrary}, $\ell_{\infty}/\ell_2$ gives already the desired result. So, we focus on the case of $k/\epsilon = \Omega(\log^5 n)$.
We pick a hash function $h: [n] \rightarrow [B]$, where $B = ck/(\epsilon \log n)$ for some constant $c$ large enough. The following follows by an application of Bernstein's Inequality and the Chernoff Bound, similarly to $\ell_{\infty}/\ell_2$.

\begin{lemma}
With probability $1-1/\mathrm{poly}(n)$, the following holds:
\[ \forall j \in [B]: |H_{k/\epsilon}(x) \cap h^{-1}(j)| \leq \log n,\qquad\text{and}\qquad  \left| \sum_{ i \in h^{-1}(j) \setminus H_{k/\epsilon}(x) } x_i^2\right| \leq \frac{\epsilon}{k} \|x_{-k}\|_2^2.	\]
\end{lemma}

We now run the $\ell_{\infty}/\ell_2$ algorithm for the previous section on vectors $x_{h^{-1}(1)},x_{h^{-1}(2)}, \ldots, x_{h^{-1}(B)}$ with sparsity parameter $\Oh(\log n)$, to obtain vectors $\hat{x}_1,\hat{x}_2, \ldots, \hat{x}_B$. The number of rounds is $\Oh(\loglog (n))$, since we can run the algorithm in every bucket in parallel. By the definition of the $\ell_{\infty}/\ell_2$ algorithm, one can see that $|\mathrm{supp}(\hat{x}_j)| \leq \Oh(\log n)$. We set $ S = \cup_{j \in B} |\mathrm{supp}(x_j)|$, and observe that $ |S| = ck/(\epsilon \log n) \cdot \Oh( \log n)   = \Oh(k/\epsilon) $. The number of measurements equals $ ck/(\epsilon \log n) \cdot \Oh( \log n \cdot \loglog (n\log (n/k)))= \Oh((k/\epsilon) \cdot \loglog (n\log (n/k)))$.

\begin{lemma}

With probability $1-1/\mathrm{poly}(n)$, we have that $ |S \setminus H_{k/\epsilon}(x)| \leq \frac{k}{\epsilon \log^2 n}.$

\end{lemma}
\begin{proof}

Since every call to $\ell_{\infty}/\ell_2$ fails with probability $1/\mathrm{poly}(\log n)$, the probability that we have more than a $\frac{1}{\log n}$ fraction of the calls that fail is at most
$${B  \choose B/\log^2 n  } \left(\frac{1}{\log n}\right)^{B/\log n } \leq (e\log^2n)^{\log n} (\log n)^{-B/\log n} \leq \frac{1}{\mathrm{poly}(n)}.$$
This implies that $S$ will contain all but at most $B/\log^2 n \cdot \log n = k/ (\epsilon \log^2 n)$ coordinates $i \in H_k(x)$.

\end{proof}

We now observe $x_S$ directly and form the vector $z = x - x_S$, for which $\|z_{-k/(\epsilon \log^2 n)}\|_2 \leq \|x_{-k/\epsilon}\|_2$. We now run a standard $\ell_2/\ell_2$ algorithm that fails with probability $1/\mathrm{poly}(n)$ to obtain a vector $\hat{z}$ that approximates $z$ (for example $\textsc{PartitionCountSketch}(z,k/(\epsilon \log^2 n), [n])$ suffices). We then output $\hat{z}+x_S$, for which
	$\|\hat{z}+x_S -x \|_2 = \|\hat{z} - z\|	\leq (1+\epsilon) \|z_{-k/(\epsilon \log n)}\|_2 \leq (1+\epsilon) \|x_{-k}\|_2.$
The number of measurements of this step is $\Oh(\frac{1}{\epsilon}\frac{k}{\log^2 n} \cdot \log n) = o( \frac{k}{\epsilon})$. The total number of rounds is clearly $\Oh(\loglog (n \log(\frac{n \epsilon}{k})))$.

\section{$\ell_2/\ell_2$ with Improved Dependence on $\epsilon$}

In this section, we prove the second part of Theorem \ref{theorem: l2l2 improved}. We first need an improved algorithm for the $1$-sparse recovery problem. 

 







\begin{lemma}\label{lem:improvedonesparse}
Let $x \in \mathbb{R}^n$. There exists an algorithm $\textsc{ImprovedOneSparseRecovery}$, that uses $\Oh(\loglog n + \frac{1}{\epsilon} \loglog (\frac{1}{\epsilon}))$ measurements in $\Oh(\loglog (n))$ rounds, and finds with sufficiently small constant probability an $\Oh(1)$-sparse vector $\hat{x}$ such that $\|\hat{x} - x\|_2 \leq (1+\epsilon) \|x_{-1}\|_2.$
\end{lemma}

\begin{proof}

We pick a hash function $h:[n] \rightarrow [B]$, where $B = \lceil 1/\epsilon^h \rceil$ for a sufficiently large constant $h$. Observe that all elements of $H_{\sqrt{B}}(x)$ are perfectly hashed under $h$ with constant probability, and, $\forall j \in [B],$ $\mathbb{E} \left[ \left\|x_{h^{-1}(j) \setminus H_{\sqrt{B}}}(x)\right\|_2 \right] \leq 1/B \|x_{-\sqrt{B}}\|_2.$
As in the previous sections, invoking Bernstein's inequality we can get that with probability $ 1 - 1/\mathrm{poly}(B)$, $\forall j \in [B],$ $\left\| x_{h^{-1}(j) \setminus H_{\sqrt{B}}(x)\|_2} \right\|_2^2 \leq \frac{c\log B}{B} \|x_{-\sqrt{B}}\|_2^2,$
where $c$ is some absolute constant, and the exponent in the failure probability is a function of $c$.

We now define the vector $z \in \mathbb{R}^{B}$, the $j$-th coordinate of which equals $z_j = \sum_{i \in h^{-1}(j)} \sigma_{i,j} x_i.$
We shall invoke Khintchine inequality to obtain $\forall j$, 
$\Pr \left[  \left|\sum_{i \in h^{-1}(j) \setminus H_{\sqrt{B}}(x)} \sigma_{i,j}x_i\right|^2 > \frac{c'}{\epsilon} \left\|x_{h^{-1}(j) \setminus H_{\sqrt{B}}(x)}\right\|_2^2 \right] \leq e^{-\Omega(1/\epsilon^2)}$,
for some absolute constant $c'$.
This allows us to take a union-bound over all $B = \lceil 1/\epsilon^h \rceil $ entries of $z$ to conclude that there exists an absolute constant $\zeta$ such that $\forall j \in [B],$ $\left|\sum_{i \in h^{-1}(j) \setminus H_{\sqrt{B}}(x)} \sigma_{i,j}x_i \right|^2 \leq \frac{c'}{\epsilon} \|x_{h^{-1}(j) \setminus H_{\sqrt{B}}(x)}\|_2^2 < \zeta \epsilon \|x_{-1}\|_2^2,$
by setting $h$ large enough.
Now, for every coordinate $j \in [B]$ for which $h^{-1}(j) \cap H_{1,\epsilon}(x) = i^*$ or some $i^* \in [n]$, we have that 
$\left| z_j \right| \geq \left|\left|x_{i^*}\right| -  \sqrt{\frac{c\log B}{B} \cdot \frac{c'}{\epsilon}} \|x_{-\sqrt{B}}\|_2  \right|	\geq (1-\zeta) \sqrt{\epsilon} \|x_{-1}\|_2$, 
whereas for every $j \in [B]$ such that $h^{-1}(j) \cap H_{1, \epsilon \zeta }(x) = \emptyset $ it holds that $|z_j| \leq 2\zeta \sqrt{\epsilon} \|x_{-1}\|_2$. We note that $H_{1,\epsilon}(x) \subset H_{\sqrt{B}}(x)$, and hence all elements of $H_{1,\epsilon}(x)$ are also perfectly hashed under $h$.
Moreover, observe that $\mathbb{E}\|z_{-1}\|_2^2 \leq \|x_{-1}\|_2^2$, and hence by Markov's inequality, we have that $\|z_{-1}\|_2^2 \leq 10 \|x_{-1}\|_2^2$ holds with probability $9/10$.
We run the $\ell_2/\ell_2$ algorithm of Theorem \ref{theorem: l2l2 improved} for vector $z$ with the sparsity being set to $1$, and obtain vector $\hat{z}$. We then set $S = \mathrm{supp}(\hat{z})$. We now define $w = (|z_1|,|z_2|,\ldots)$, for which $\|w_{-1}\|_2 = \|z_{-1}\|_2$.
Clearly,
$\|z - z_S \|_2^2 \leq \|z-\hat{z}\|_2^2 \leq (1+\epsilon) \|z_{-1}\|_2^2 = (1+\epsilon) \|w_{-1}\|_2^2$. 
So $\|w - w_S \|_2^2 = \|z-z_S\|_2^2 \leq (1+\epsilon) \|w_{-1}\|_2^2$. 
We now prove that
$\left\|x - x_{\cup_{j \in S} h^{-1}(j)}\right\|_2 \leq (1+\Oh(\epsilon))\|x_{-1}\|_2$. 
Let $i^*$ be the largest coordinate in magnitude of $x$, and $j^* = h(i^*)$. If $j^* \in S$, then it follows easily that $\|x - x_{\cup_{j \in S} h^{-1}(j)}\|_2 \leq \|x_{-1}\|_2$. Otherwise, since $\sum_{j \neq j^*} w_j^2 = \|w_{-1}\|_2^2$, and $\sum_{j \notin S} w_j^2 \leq (1+\epsilon)\|w_{-1}\|_2^2$, it must be the case that 
$\left|w_{j^*}^2 - \|w_S\|_2^2\right| \leq \epsilon \|w_{-1}\|_2^2 \leq 10\epsilon \|x_{-1}\|_2^2$. 
The above inequality, translates to
$\sum_{i \in h^{-1}(j^*)}x_i^2 \leq |S|  \zeta \epsilon \|x_{-1}\|_2^2 + \zeta \epsilon \|x_{-1}\|_2^2 + 10\epsilon \|x_{-1}\|_2^2 + \sum_{j \in S}\sum_{i \in h^{-1}(j)} x_j^2  = \Oh(\epsilon) \|x_{-1}\|_2^2 + \sum_{j \in S}\sum_{i \in h^{-1}(j)} x_j^2$. 
This gives  
$\left\|x - x_{\cup_{j \in S} h^{-1}(j)}\right\|_2 = \sum_{i \in h^{-1}(j^*)}x_i^2 + \sum_{j \notin S\cup\{j^*\}} \sum_{i \in h^{-1}(j)} x_i^2 \leq  \Oh(\epsilon)\|x_{-1}\|_2^2 + \Oh(1) \zeta\epsilon \|x_{-1}\|_2^2 + \sum_{j \in S}\sum_{i \in h^{-1}(j)} x_j^2 + \sum_{j \notin S\cup\{j^*\}} \sum_{i \in h^{-1}(j)} x_i^2 + \leq (1+\Oh(\epsilon)) \|x_{-1}\|_2^2.$	






Given $S$, we run the $1$-sparse recovery routine on vectors $x_j$ for $j \in S$, with a total of $\Oh(\loglog n)$ measurements and $\Oh(\loglog n)$ rounds. We then output $\{x_{i_j}\}_{j \in S}$. Let $i_j$ be the index returned for $j \in S$ by the $1$-sparse recovery routine. Since we have a constant number of calls to the $1$-sparse recovery routine (because $S$ is of constant size), all our $1$-sparse recovery routines will succeed. We now have that
$\|x - x_{ \cup_{j \in S} i_j} \|_2 \leq \|x_{\bar{S}}\|_2 + \sum_{j \in S} \|x_{h^{-1}(j)} - x_{i_j}\|_2
\leq \|x_{\bar{S}}\|_2  + \sum_{j \in S} (1+\epsilon) \|x_{h^{-1}(j) \setminus H_1(x)} \|_1 \leq (1+\Oh (\epsilon )) \|x_{-1}\|_2.$
Rescaling $\epsilon$, we get the desired result.
\end{proof}

The algorithm for general $k$ is similar to \cite{indyk2011power}, apart from the fact that we subsample at a slower rate, and also use our new $1$-sparse recovery algorithm as a building block. In the algorithm below, $R_r$ is the universe we are restricting our attention on at the $r$th round. Moreover, $J$ is the set of coordinates that we have detected so far. We are now ready to prove Theorem \ref{theorem: l2l2 improved}.

\begin{algorithm}
\label{l1/l1a}
\caption{Adaptive $\ell_2/\ell_2$ Sparse Recovery}
\label{algorithm: ell2}
\begin{algorithmic}
\STATE {\bfseries 1.} $R_0 \leftarrow [n]$.
\STATE {\bfseries 2.} $x_0 \leftarrow \vec{0}$.
\STATE {\bfseries 3.} $\delta_0 \leftarrow \delta/2, \epsilon_0 \leftarrow \epsilon/ e,f_0 \leftarrow 1/32,k_0 \leftarrow k$.
\STATE {\bfseries 4.} $J \leftarrow \emptyset$.
\STATE {\bfseries 5.} \textbf{For} $r=0$ to $\Oh(\log^*k)$ \textbf{do}
\STATE {\bfseries 6.} \qquad \textbf{For} $t=0$ to $\Theta(k_r\log(1/(\delta_rf_r)))$ \textbf{do}
\STATE {\bfseries 7.} \qquad \qquad $S_t \leftarrow \textsc{Subsample}(x - x^{(r)},R_r,1/(C_0k_r))$.
\STATE {\bfseries 8.} \qquad \qquad $J \leftarrow J \cup \textsc{ImprovedOneSparseRecovery}((x-x^{(r)})_{S_t})$.
\STATE {\bfseries 9.} \qquad \textbf{End For}
\STATE {\bfseries 10.} \quad\ \ $R_{r+1} \leftarrow [n] \setminus J$.
\STATE {\bfseries 11.} \quad\ \ $\delta_{r+1} \leftarrow \delta_r/8$.
\STATE {\bfseries 12.} \quad\ \ $\epsilon_{r+1} \leftarrow \epsilon_r/2$.
\STATE {\bfseries 13.} \quad\ \ $f_{r+1} \leftarrow 1/2^{1/(4^{i+r}f_r)}$.
\STATE {\bfseries 14.} \quad\ \ $k_{r+1} \leftarrow f_r k_r$.
\STATE {\bfseries 15.} \quad\ \ $R_{r+1} \leftarrow [n] \setminus J$.
\STATE {\bfseries 16.} \textbf{End For}
\STATE {\bfseries 17.} $\hat{x} \leftarrow x^{(r+1)}$.
\STATE {\bfseries 18.} Return $\hat{x}$.
\end{algorithmic}
\end{algorithm}

\begin{proof}
The number of measurements is bounded in the exact same way as in Theorem 3.7 from \cite{indyk2011power}.

We fix a round $r$ and $i \in H_{k_r,\epsilon_r}(x^{(r)})$. Then the call to $\textsc{Subsample}(R_r,1/(C_0k_r))$ yields
\[	\Pr \left[ |H_{k_r,\epsilon_r}(x - x^{(r)}) \cap S_t| = \{i\} \right] \geq \frac{1}{C_0k_r},\quad	\mathbb{E} \left[ \|x_{S_t \setminus H_{k_r,\epsilon_i}(x^{(r)})} \|_2^2 \right] = \frac{1}{C_0 k_r}\|x_{-k_r}\|_2^2. \]
Setting $C_0$ to be large enough and combining Markov's inequality with the guarantee of Lemma \ref{lem:improvedonesparse}, we get that the probability that the call to $\textsc{ImprovedOneSparseRecovery}(x_{S_t})$ returns $i$ is $\Theta(1/k_r)$. Because we repeat $k_r \log(1/(f_r \delta_r))$, the probability that $i$ or a set $S_i$ of size $\Oh(1)$ such that $\|x_{\{i\}} - x_{S_i}\|_2 \leq \epsilon_i \|x_{-k_r}\|_2^2$, is not added in $J$ is at most $(1-1/k_r)^{k_r \log(1/(f_r\delta_r))} = f_r\delta_r$.

Given the above claim, the number of measurements is $\Oh((k \loglog n + k/ \epsilon \loglog (1/\epsilon) \log(1/\delta))$ and the analysis of the iterative loop proceeds almost identically to Theorem 3.7 of \cite{indyk2011power}. 

  \end{proof}

\bibliographystyle{alpha}
\bibliography{reference}

\newpage
\appendix
\section{Toolkit}

\begin{lemma}[Bernstein's Inequality]
There exists an absolute constant $c_B$ such that for independent random variables $X_1,\ldots,X_r,$ with $|X_i| \leq K$ we have that
\[	\forall \lambda>0, \Pr\left[ \left|\sum_i X_i - \mathbb{E} \sum_i X_i \right| > \lambda \right] \leq e^{-C_B \lambda/\sigma^2} + e^{-C_B \lambda/K},		\]
where $\sigma^2 = \sum_i \mathbb{E}(X_i - \mathbb{E}X_i)^2$.

\end{lemma}

\begin{lemma}[\cite{li2017low}] \label{onesparse}
Let $x \in \mathbb{R}^n$ be such that there exists a coordinate $i$ for which $|x_i| \geq 5 \|x_{-k}\|_2$. There exists an adaptive algorithm $\textsc{OneSparseRecovery}(x)$ that uses $\Oh(\log \log n)$ measurements and $\Oh( \log \log n)$ rounds and finds $i$ with probability $1 - 1/\mathrm{poly}(\log n)$.
\end{lemma}

\begin{lemma}[\textsc{PartitionCountSketch} \cite{larsen2016heavy}]
Let $z \in \mathbb{R}^n$ and let $\mathcal{P} =\{P_1,P_2,\ldots,P_{|\mathcal{P}|}\}$ be a partition of $[n]$. Then, there exists a non-adaptive scheme $\textsc{PartitionCountSketch}(x,k,\mathcal{P})$ that uses
$\Oh( k \log(|\mathcal{P}|))$ measurements, which computes a vector $w \in \mathbb{R}^{|\mathcal{P}|}$ such that 	
\[ \forall j \in [|\mathcal{P}|]:  \|w_{j}\|_2^2 \in \left[ (1-\gamma) \|x_{P_j}\|_2^2 - \frac{1}{k} \|x_{-k}\|_2^2,\ (1+\gamma) \|x_{P_j}\|_2^2 + \frac{1}{k} \|x_{-k}\|_2^2 \right] ,\]
where $\gamma$ is an arbitrary small constant. The failure probability of the scheme is $\frac{1}{\mathrm{poly}(|\mathcal{P}|)}$.

\end{lemma}

\section{Adaptive $\ell_p/\ell_p$ Recovery Lower Bounds}
%
%
%

%
In Section
\ref{sec:cc1} we briefly introduce the definition and lower bounds for the communication complexity
of $\indlinf$,  a two-party communication problem that is defined and studied in \cite{price20111+}.
In Section \ref{sec:cc2} we show how to use an adaptive $(1+\eps)$-approximate $\ell_p/\ell_p$ sparse
recovery scheme $\mathcal{A}$
to solve the communication problem $\indlinf$. By the communication lower bound in Section \ref{sec:cc1},
we obtain a lower bound on the number of measurements required of an adaptive $(1+\eps)$-approximate $\ell_p/\ell_p$ sparse
recovery scheme.

\subsection{Direct Sum for Distributional $\ell_{\infty}$}\label{sec:cc1}
Consider the two-party randomized communication complexity setting. 
There are two parties,
Alice and Bob, with input vectors $x$ and $y$ respectively, and their goal is
to solve a promise problem $f(x,y)$. The parties have private randomness.
The communication cost of a protocol is
its maximum transcript length, over all possible inputs and random coin tosses.
The randomized communication complexity $R_{\delta}(f)$ is
the minimum communication cost of a randomized protocol $\Pi$ which for every input
$(x,y)$, outputs $f(x,y)$ with probability at least $1-\delta$ (over the random
coin tosses of the parties). We also study the distributional
complexity of $f$, in which the parties are deterministic and
the inputs $(x,y)$ are drawn from distribution $\mu$, and a protocol is
correct if it succeeds with probability at least $1-\delta$ in outputting $f(x,y)$,
where the probability is now taken over $(x,y) \sim \mu$. We define $D_{\mu, \delta}(f)$
to be the minimum communication cost of a correct protocol $\Pi$.

We consider the following promise problem
$\gaplinf^B$, where $B$ is a parameter,
which was studied in \cite{saks2002space,bar2004information}. The
inputs are pairs $(x,y)$ of $m$-dimensional vectors, with
$x_i, y_i \in \{0, 1, 2, \ldots, B\}$ for all $i \in [m]$, with the promise that
$(x,y)$ is one of the following types of instances:
\begin{itemize}
\item NO instance: for all $i$, $|x_i-y_i| \in \{0,1\}$, or
\item YES instance: there is a unique $i$ for which
  $|x_i-y_i| = B$, and for all $j \neq i$, $|x_j-y_j| \in \{0,1\}$.
\end{itemize}
The goal of a protocol is to decide which of the two cases (NO or YES) the input
is in.
Consider the distribution $\sigma$: for each $j \in
[m]$, choose a random pair $(Z_j, P_j) \in \{0, 1, 2, \ldots, B\} \times \{0,1\}
\setminus \{(0,1), (B, 0)\}$. If $(Z_j, P_j) = (z, 0)$, then $X_j = z$ and
$Y_j$ is uniformly distributed in $\{z, z+1\}$; if $(Z_j, P_j) = (z, 1)$, then $Y_j = z$
and $X_j$ is uniformly distributed on $\{z-1, z\}$.
Let $Z = (Z_1, \ldots, Z_m)$ and $P = (P_1, \ldots, P_m)$.
Next choose a random coordinate $S \in [m]$.
For coordinate $S$, replace $(X_{S}, Y_{S})$ with a uniform element of
$\{(0,0), (0,B)\}$.  Let $X = (X_1, \ldots, X_m)$ and $Y =
(Y_1, \ldots, Y_m)$.
In \cite{price20111+}, a problem, $\indlinf^{r, B}$ is defined,  which involves solving $r$ copies of $\gaplinf^B$. This is related to the $\ell_1/\ell_1$ recovery scheme with  $\indlinf^{r, B}$ in order to obtain a lower bound. Here we introduce the definition of  $\indlinf^{r, B}$ and first present their results.
%
\begin{definition}[Indexed $\indlinf^{r,B}$ Problem] There are
  $r$ pairs of inputs $(x^1,y^1),(x^2,y^2),\ldots,(x^r,y^r)$ such that
  every pair $(x^i,y^i)$ is a legal instance of the $\gaplinf^B$
  problem. Alice is given $x^1, \ldots, x^{r}$.  Bob is given
  an index $I \in [r]$ and $y^1, \ldots, y^r$.  The goal is to
  decide whether $(x^I, y^I)$ is a NO or a YES instance of
  $\gaplinf^B$.
\end{definition}

Let $\eta$ be the distribution $\sigma^r \times U_r$, where $U_r$ is the
uniform distribution on $[r]$. We bound
$D^{1-way}_{\eta, \delta}(\indlinf)^{r,B}$ as follows.
For a function $f$, let $f^r$ denote the problem of computing $r$ instances of $f$.
For a distribution $\zeta$ on instances of $f$,
let $D_{\zeta^r, \delta}^{1-way, *} (f^r)$ denote the minimum communication cost of a deterministic
protocol computing a function $f$ with error probability at most $\delta$ {\it in each}
of the $r$ copies of $f$, where the inputs come from $\zeta^r$.

\begin{theorem}\label{thm:theMain}
For $\delta$ less than a sufficiently small constant,
$D^{1-way}_{\eta, \delta}(\indlinf^{r,B}) = \Omega(\delta^2 r m/(B^2 \log r))$.
\end{theorem}

\begin{lemma}\label{lemma:l1entropy}
  Let $R = [s, cs]$ for some constant $c$ and parameter $s$.  Let $X$
  be a permutation-independent distribution over $\{0, 1\}^{n}$ with
  $\norm{1}{x} \in R$ with probability $p$.  If $y$ satisfies
  $\norm{1}{x-y} \leq (1-\eps)\norm{1}{x}$ with probability $p'$ with
  $p' - (1-p) = \Omega(1)$, then $I(x; y) = \Omega(\eps s\log (n/s))$.
\end{lemma}

\begin{lemma}\label{lem:bits}
  A lower bound of $\Omega(b)$ bits for such an adaptive $\ell_p/\ell_p$ sparse recovery bit
  scheme with $0<p \leq 2$ implies a lower bound of
  $\Omega(b/((1+c+d)\log n))$ bits for $(1+\eps)$-approximate
  sparse recovery with failure probability $\delta - 1/n$.
\end{lemma}
\subsection{The Overall Lower Bound}\label{sec:cc2}
%

The proof of the adaptive lower bound for $\ell_p/\ell_p$ schemes is similar to the proof for the non-adaptive lower bound for $\ell_1/\ell_1$
sparse recovery given in \cite{price20111+}. 
Fix parameters $B = \Theta(1/\eps^{1/2})$, $r = k$, $m =
1/\eps^{(2+p)/2}$, and $n = k/\eps^3$. Given an instance $(x^1, y^1),
\ldots, (x^r, y^r)$ of  $\indlinf^{r, B}$ we define the input signal
$z$ to a sparse recovery problem. We allocate a set $S^i$ of $m$
disjoint coordinates in a universe of size $n$ for each pair $(x^i,
y^i)$, and on these coordinates place the vector $y^i-x^i$.  The
locations turn out to be essential for the proof of Lemma \ref{lem:heavyHitters} below,
and are placed uniformly at random among the $n$ total coordinates (subject to the
constraint that the $S^i$ are disjoint).
Let $\rho$ be the induced distribution on $z$.

Fix an $\ell_p/\ell_p$ recovery multiround bit scheme $\mathcal{A}$ that
uses $b$ bits and succeeds with
probability at least $1-\delta_1/2$ over $z \sim \rho$.
Let $S$ be the set of top
$k$ coordinates in $z$. As shown in equation (14) of \cite{price20111+},
$\mathcal{A}$ has the guarantee that if $v = \mathcal{A}(z)$, then
\begin{eqnarray}\label{eqn:main}
\|(v-z)_S\|_p^p + \|(v-z)_{[n] \setminus S}\|_p^p \leq (1+2\eps)\|z_{[n] \setminus S}\|_p^p.
\end{eqnarray}
Next is our generalization of Lemma 6.8 of \cite{price20111+}.

\begin{lemma}\label{lem:heavyHitters}
  For $B= \Theta(1/\eps^{1/2})$ sufficiently large, suppose that
  $\Pr_{z \sim \rho}[\|(v-z)_S\|_p^p \leq 10 \eps \cdot \|z_{[n]
    \setminus S}\|_p^p] \geq 1-\delta$.  Then $\mathcal{A}$ requires $b =
  \Omega(k/(\eps^{p/2} \log k))$.
\end{lemma}
\begin{proof}
  We need to show how to use $\mathcal{A}$ to solve instances of $\indlinf^{r,B}$ with
  probability at least $1- C$ for some small $C$, where the
  probability is over input instances to $\indlinf^{r,B}$ distributed
  according to $\eta$, inducing the distribution $\rho$. 
  Since $\mathcal{A}$ is a
  deterministic sparse recovery bit scheme, it receives a sketch
  $f(z)$ of the input signal $z$ and runs an arbitrary recovery
  algorithm $g$ on $f(z)$ to determine its output $v = \mathcal{A}(z)$.

  Given $x^1, \ldots, x^r$, for each $i = 1, 2, \ldots, r$, Alice places
  $-x^i$ on the appropriate coordinates in the block $S^i$ used in
  defining $z$, obtaining a vector $z_{Alice}$, and transmits
  $f(z_{Alice})$ to Bob.  Bob uses his inputs $y^1, \ldots, y^r$ to
  place $y^i$ on the appropriate coordinate in $S^i$. He thus creates a
  vector $z_{Bob}$ for which $z_{Alice} + z_{Bob} = z$. Given
  $f(z_{Alice})$, Bob computes $f(z)$ from $f(z_{Alice})$ and
  $f(z_{Bob})$, then $v = \mathcal{A}(z)$. We assume all coordinates of $v$
  are rounded to the real interval $[0, B]$, as this can only decrease
  the error.

  We say that $S^i$ is {\it bad} if either
  \begin{itemize}
  \item there is no coordinate $j$ in $S^i$ for which $|v_j| \geq
    \frac{B}{2}$ yet $(x^i, y^i)$ is a YES instance of
    $\gaplinf^{r,B}$, or
  \item there is a coordinate $j$ in $S^i$ for which $|v_j| \geq
    \frac{B}{2}$ yet either $(x^i, y^i)$ is a NO instance of
    $\gaplinf^{r, B}$ or $j$ is not the unique $j^*$ for which $y^i_{j^*} - x^i_{j^*} = B$
  \end{itemize}
  For $B$ sufficiently large, the $\ell_p$-error incurred by a bad block is at least
  $B/4$. Hence, if there are $t$ bad blocks, the total error to the $p$-th power is at
  least $t B^p/4^p$, which must be smaller than $10 \eps \cdot \|z_{[n]
    \setminus S}\|_p^p$ with probability $1-\delta$. Conditioned on this, we would like to bound $t$.  All coordinates in $z_{[n] \setminus S}$ have value
  in the set $\{0, 1\}$. Hence, $\|z_{[n] \setminus S}\|_p^p \leq rm$.  So
  $t \leq 4^p10 \eps rm /B^p\leq 160 \eps rm/B^p$.
  Plugging in $r$, $m$ and $B$, 
  $t \leq C k$, where $C > 0$ is a constant that can be made
  arbitrarily small by increasing $B = \Theta(1/\eps^{1/2})$.

  If a block $S^i$ is not bad, then it can be used
  to solve $\gaplinf^{r, B}$ on $(x^i, y^i)$ with probability
  $1$. Bob declares that $(x^i, y^i)$ is a YES instance if and
  only if there is a coordinate $j$ in $S^i$ for which $|v_j| \geq
  B/2$.

  Since Bob's index $I$ is uniform on the $m$ coordinates in
  $\indlinf^{r,B}$, with probability at least $1-C$ the players solve
  $\indlinf^{r, B}$ given that the $\ell_p$ error is small. Therefore
  they solve $\indlinf^{r, B}$ with probability $1-\delta-C$
  overall. By Theorem \ref{thm:theMain}, for $C$ and $\delta$
  sufficiently small, $\mathcal{A}$ requires $\Omega(mr/(B^2 \log r)) =
  \Omega(k/(\eps^{p/2} \log k))$ bits.
\end{proof}

\begin{lemma}\label{lem:entropy}
  Suppose $\Pr_{z \sim \rho}[\|(v-z)_{[n] \setminus S}\|_p^p] \leq
  (1-8\eps)\cdot \|z_{[n] \setminus S}\|_p^p] \geq \delta/2$.  Then
  $\mathcal{A}$ requires $b = \Omega(\frac{1}{\eps^{p/2}}k \log
  (1/\eps))$.
\end{lemma}
\begin{proof}
  The distribution $\rho$ consists of $B(mr, 1/2)$ ones placed
  uniformly throughout the $n$ coordinates, where $B(mr, 1/2)$ denotes
  the binomial distribution with $mr$ events of $1/2$ probability each.
  Therefore with probability at least $1-\delta/4$, the number of ones
  lies in $[\delta mr / 8, (1 - \delta/8)mr]$.  Thus by
  Lemma~\ref{lemma:l1entropy}, $I(v; z) \geq \Omega(\eps mr \log
  (n/(mr)))$.  Since the mutual information only passes through a $b$-bit
  string, $b = \Omega(\eps mr \log (n/(mr))) = \Omega(\frac{1}{\eps^{p/2}}k \log
  (1/\eps))$ as well.
\end{proof}

\begin{theorem}
  Any adaptive $(1+\eps)$-approximate $\ell_p/\ell_p$ recovery scheme with
  sufficiently small constant failure probability $\delta$ must make
  $\Omega(\frac{1}{{\eps}^{p/2}}k/\log^2 (k/\eps))$ measurements.
\end{theorem}
\begin{proof}
  We will lower bound any $\ell_p/\ell_p$ sparse recovery bit scheme
  $\mathcal{A}$. If $\mathcal{A}$ succeeds, then in order to satisfy inequality
  (\ref{eqn:main}), we must either have $\|(v-z)_S\|_p^p \leq 10 \eps
   \|z_{[n] \setminus S}\|_p^p$ or we must have $\|(v-z)_{[n]
    \setminus S}\|_p^p \leq (1-8\eps) \|z_{[n] \setminus
    S}\|_p^p$. Since $\mathcal{A}$ succeeds with probability at least
  $1-\delta$, it must either satisfy the hypothesis of Lemma
  \ref{lem:heavyHitters} or the hypothesis of Lemma
  \ref{lem:entropy}. But by these two lemmas, it follows that $b =
  \Omega(\frac{1}{{\eps}^{p/2}}k/\log k)$.  Therefore by
  Lemma~\ref{lem:bits}, any $(1+\eps)$-approximate
  $\ell_p/\ell_p$ sparse recovery algorithm requires
  $\Omega(\frac{1}{{\eps}^{p/2}}k/\log^2 (k/\eps))$ measurements.
\end{proof}

\end{document}